\documentclass[twocolumn,showpacs,%
  aps,superscriptaddress,%
  eqsecnum,prd,notitlepage,showkeys,
  footinbib]{revtex4-1}

\usepackage{amssymb}
\usepackage{amsmath}
\usepackage{graphicx}
\usepackage{dcolumn}
\usepackage[colorlinks,urlcolor=blue,citecolor=blue,linkcolor=blue]{hyperref}
\usepackage{lipsum}
\usepackage{color,units}
\usepackage[dvipsnames]{xcolor} 
\usepackage{aas_macros}
\usepackage{lineno}
\usepackage{xspace}

\usepackage{siunitx}
\DeclareSIUnit \parsec {pc}

\usepackage[normalem]{ulem} 

\newcommand{\nrsur}{\textsc{NRHybSur3dq8}\xspace}
\newcommand{\imr}{\textsc{IMRPhenomD}\xspace}

\newcommand{\BF}{\text{BF}}

\newcommand{\SPA}{School of Physics and Astronomy, Monash University, Clayton, VIC 3800, Australia}
\newcommand{\OzGravMonash}{OzGrav: The ARC Centre of Excellence for Gravitational Wave Discovery, Clayton, VIC 3800, Australia}

\begin{document}

\title{Thanks for the memory: measuring gravitational-wave memory\\ in the first LIGO/Virgo gravitational-wave transient catalog}

\author{Moritz H\"ubner}
    \email{moritz.huebner@monash.edu}
\author{Colm Talbot}
    \email{colm.talbot@monash.edu}
\author{Paul D. Lasky}
    \email{paul.lasky@monash.edu}
\author{Eric Thrane}
    \email{eric.thrane@monash.edu}
\affiliation{\SPA}
\affiliation{\OzGravMonash}

\begin{abstract}
Gravitational-wave memory, a strong-field effect of general relativity, manifests itself as a permanent displacement in spacetime.
We develop a Bayesian framework to detect gravitational-wave memory with the Advanced LIGO/Virgo detector network.
We apply this algorithm on the ten binary black hole mergers in LIGO/Virgo's first transient gravitational-wave catalog.
We find no evidence of memory, which is consistent with expectations.
In order to estimate when memory will be detected, we use the best current population estimates to construct a realistic sample of binary black hole observations for LIGO/Virgo at design sensitivity.
We show that an ensemble of $\mathcal{O}(2000)$ binary black hole observations can be used to find definitive evidence for gravitational-wave memory.
We conclude that memory is likely to be detected in the early A+/Virgo+ era.
\end{abstract}

\maketitle


\section{Introduction}\label{sec:introduction}
Gravitational waves from binary black hole mergers are now observed regularly with LIGO and Virgo~\cite{aLIGO, aVirgo, GWTC1}.
These observations allow us to investigate aspects of general relativity that could not have been studied observationally until now~\cite{GW150914TestingGR, GW170817TestingGR, GWTC1TestingGR, Ghosh2016}.
One such aspect is gravitational-wave memory, a strong-field effect of general relativity that is sourced from the emission of gravitational waves.
Memory causes a permanent displacement between freely falling test masses~\cite{ZeldovichYa.1974, Braginsky1987, Thorne1992}.

In general, memory can arise both in the linearized Einstein field equations and in their full non-linear form.
Early research focused on the production of linear memory from unbound systems such as supernovae or triple black hole interactions~\cite{Braginsky1987}.
Non-linear contributions to memory were originally thought to be negligibly small~\cite{DemetriosChristodoulou1991}.
However, further investigations showed that bound systems such as binary black holes produce significant non-linear memory~\cite{DemetriosChristodoulou1991, Thorne1992}.
Non-linear memory can be interpreted as the component of a gravitational wave that is sourced by the emission of the gravitational wave itself~\cite{Thorne1992}.
The amplitude of memory is typically no more than $\mathcal{O}(5 \%)$ of the peak oscillatory waveform amplitude in typical binary black hole systems~\cite{Braginsky1987}.
Detecting gravitational-wave memory from a single merger with current generation detectors is improbable due to the low amplitude of memory~\cite{Lasky2016, Johnson2018}.

Memory will be detectable from single events with proposed future detectors such as LISA, Cosmic Explorer, and the Einstein Telescope~\cite{FutureDetectorsAbbott2017, ET2010, Islo2019}.
While detecting memory with LIGO/Virgo~\cite{AdvancedLigo2015, AdvancedVirgo2014} directly from a single merger is not possible, it is potentially detectable using an ensemble of mergers~\cite{Lasky2016}.
Proposed low-frequency improvements to LIGO could substantially increase the sensitivity to the memory effect~\cite{LIGO_LOW_FREQ}.
Searches for memory from supermassive black-hole binaries with pulsar timing arrays also have been proposed~\cite{VanHaasteren2010, CORDES2012, 2014Madison} and carried out (e.g.~\cite{MEMORY_PTA_PARKES2015, 2015NanoGRAV, 2019NanoGRAV}), although without any detection yet.
Future pulsar timing arrays, using data from the Square Kilometer Array~\cite{SKA2009}, may be able to detect memory from supermassive black hole binaries~\cite{Johnson2018}.

There are a number of proposed sources of memory  besides binaries.
These include high-frequency sources outside the LIGO band such as dark matter collapse in stars~\cite{Yasunari2016}, black hole evaporation~\cite{Greene2012, Nakamura1997}, or cosmic strings~\cite{Damour2000}.
While such sources are purely conjectural, they would be able to produce memory that is detectable within the LIGO band~\cite{OrphanMemory}.

Recent work has also shown that there is a redshift enhancement in memory at cosmological distances, which will become relevant for future detectors~\cite{Bieri2016, Bieri2017}.
Other theoretical work has shown the links between the memory effect, soft gravitons, and asymptotic symmetries in general relativity, which has implications for the black hole information paradox~\cite{Strominger2016, Hawking2016, Kapec2017}.
Measurements of memory with gravitational waves may eventually prove useful studying these phenomena, though, it is not yet clear how.

In this paper, we perform the first search for gravitational-wave memory using the ten binary black hole mergers that LIGO and Virgo observed during their first two observing runs \cite{GWTC1}.
We find no evidence for memory, consistent with expectations.
However, the infrastructure developed here will be used on future observations.
We show that, using $1830^{+1730}_{-1100}$ gravitational-wave observations we will be able to accumulate enough evidence to definitively detect gravitational-wave memory.
With the memory signal firmly established, it will then be possible to characterize the properties of memory to see if they are consistent with general relativity.

We structure the remainder of this paper as follows.
In Section~\ref{sec:methods}, we discuss the methods required to detect memory.
In Section~\ref{sec:results}, we apply our algorithm to the first ten binary black hole observations and report the results.
In Section~\ref{sec:population}, we use binary black hole population estimates from the first two LIGO/Virgo observing runs to create a realistic sample of future binary black-hole merger observations and calculate the required number to detect memory.
Finally, in Section~\ref{sec:conclusion} we provide an outlook for future developments.


\section{Methods}\label{sec:methods}
\subsection{Signal models}
The first major consideration in our analysis is the choice of our signal model.
The most precise signal models for binary black hole mergers are numerical-relativity simulations, which solve the Einstein field equations numerically given a set of initial conditions.
However, numerical-relativity simulations may take months to carry out even for single mergers.
Surrogate models, i.e., models that interpolate between a set of pre-computed waveforms, are hence preferred to create high fidelity waveforms in $\mathcal{O}(1\si{\second})$~\cite{Blackman2015, NRHYBSUR}.
Unfortunately, numerical-relativity waveforms and their associated surrogates typically do not include memory since memory is hard to resolve when carrying out numerical-relativity simulations~\cite{Favata2010}.

Recent advances have made it practical to calculate memory directly from the oscillatory part of the waveform~\cite{Favata2009, Favata2009a, Talbot2018}.
We use the {\sc GWMemory} package~\cite{Talbot2018}, which calculates memory from arbitrary oscillatory waveforms, which we then add to the oscillatory component to obtain the full waveform.
We compute the memory using \imr~\cite{IMRPhenomD}, a phenomenological model that describes the gravitational wave during the inspiral, merger, and ringdown phase for aligned-spin binary black holes.

One additional consideration was pointed out in~\cite{Lasky2016}.
The memory changes sign under a transformation $\phi \rightarrow \phi \pm \pi/4$ and $\psi \rightarrow \psi \pm \pi/4$.
Here $\phi$ is the phase at coalescence and $\psi$ is the polarization angle of the waveform.
At the same time, this transformation leaves the lower order spin-weighted spherical harmonic modes $(l, m) = (2, \pm 2)$ unaffected, which causes a degeneracy in the $(\phi,\psi)$ posterior space.
If we only use $(\ell, m) = (2, \pm 2)$ modes, this degeneracy implies the sign of the memory is unknown, which causes the signal to add incoherently (like the fourth root of the number of mergers).
Including higher-order modes in the signal model to break this degeneracy is hence advantageous, as they help us to determine the sign of the memory (which causes the signal to grow like the square root of the number of mergers).

\subsection{Bayesian methods}
In order to determine whether a set of gravitational-wave observations contains a memory signature, we perform Bayesian model selection using LIGO/Virgo data.
We define our ``full" signal model to be the waveform that includes both the oscillatory and memory part of the waveform.
We test this model against an ``oscillatory only" model (abbreviated ``osc'') that only contains the oscillatory part of the waveform.

The Bayes factor describes how much more likely one hypothesis is to have produced the available data compared to another.
We define the memory Bayes factor as
\begin{equation}
    \text{BF}_{\mathrm{mem}} = \frac{\mathcal{Z}_{\mathrm{full}}}{\mathcal{Z}_{\mathrm{osc}}} \, ,
\end{equation}
where $\mathcal{Z}_{\mathrm{full}}$ and $\mathcal{Z}_{\mathrm{osc}}$ are each an evidence (fully-marginalized likelihood) corresponding to our two models. 
See Ref.~\cite{Thrane2019} for a review of Bayesian statistics in the context of gravitational-wave astronomy.
The total memory Bayes factor $\BF_{\mathrm{mem}^{\mathrm{tot}}}$ can then be accumulated over a series of $N$ gravitational-wave observations,
\begin{equation}
    \BF_{\mathrm{mem}}^\text{tot} = \prod_{i=1}^{N} \BF^i_{\mathrm{mem}}\, .
\end{equation}
Following convention (e.g.~\cite{Lasky2016}), we consider $\ln \BF_{\mathrm{mem}}^{\mathrm{tot}} \geq 8$ a detection.

We calculate both the the posterior probability distributions for the model parameters and the evidence using a nested sampling algorithm~\cite{SkillingNestedSampling,FerozNestedSampling, Dynesty}.
In practice, we perform all runs in this paper using the interface to the nested-sampling package \textsc{dynesty}~\cite{Dynesty} within \textsc{Bilby}.
Stochastic sampling noise in evidence calculations dominate our results if the difference in evidence between both models is small.
We resolve this issue by sampling with the oscillatory-only model and reweighting the posterior samples to the full model to determine the Bayes factor between these two models following the prescription from~\cite{HOM_REWEIGHTING}.
A similar analysis has recently been carried out to search for eccentricity in the existing binary catalog~\cite{RomeroShaw2019}.
Given a set of $n$ posterior samples $\theta_k$ and the observed data $d$, we calculate the memory Bayes factor $\BF_{\mathrm{mem}}$ using the oscillatory-only likelihood $\mathcal{L}_{\mathrm{osc}}$ and the full likelihood $\mathcal{L}_{\mathrm{full}}$:
\begin{equation}\label{eq:reweight_1}
    \BF_{\mathrm{mem}} = \frac{1}{n}\sum_{k=1}^n \frac{\mathcal{L}_{\mathrm{\mathrm{full}}}(\theta_k | d)}{\mathcal{L}_{\mathrm{osc}}(\theta_k | d)}\, 
    \equiv \frac{1}{n}\sum_{k=1}^n w_k.
\end{equation}
We refer to the likelihood ratio $w_k$ as ``weights."
This approach is valid if both models have similar posterior distributions, which is true in our case.
Since the Bayes factor is now based on the same set of samples for both models, the stochastic sampling noise cancels.

\subsection{Reweighting study}
In order to study the performance of the reweighting technique, we simulate GW150914-like events with different signal strengths in the LIGO/Virgo detector network at design sensitivity with a zero-noise realization using the \textsc{Bilby} software package~\cite{2018Bilby}.
We create the oscillatory part of the waveform with \imr and add the memory part of the waveform by using the \textsc{GWMemory} package~\cite{Talbot2018}.

\begin{figure}
    \centering
    \includegraphics[width=0.5\textwidth]{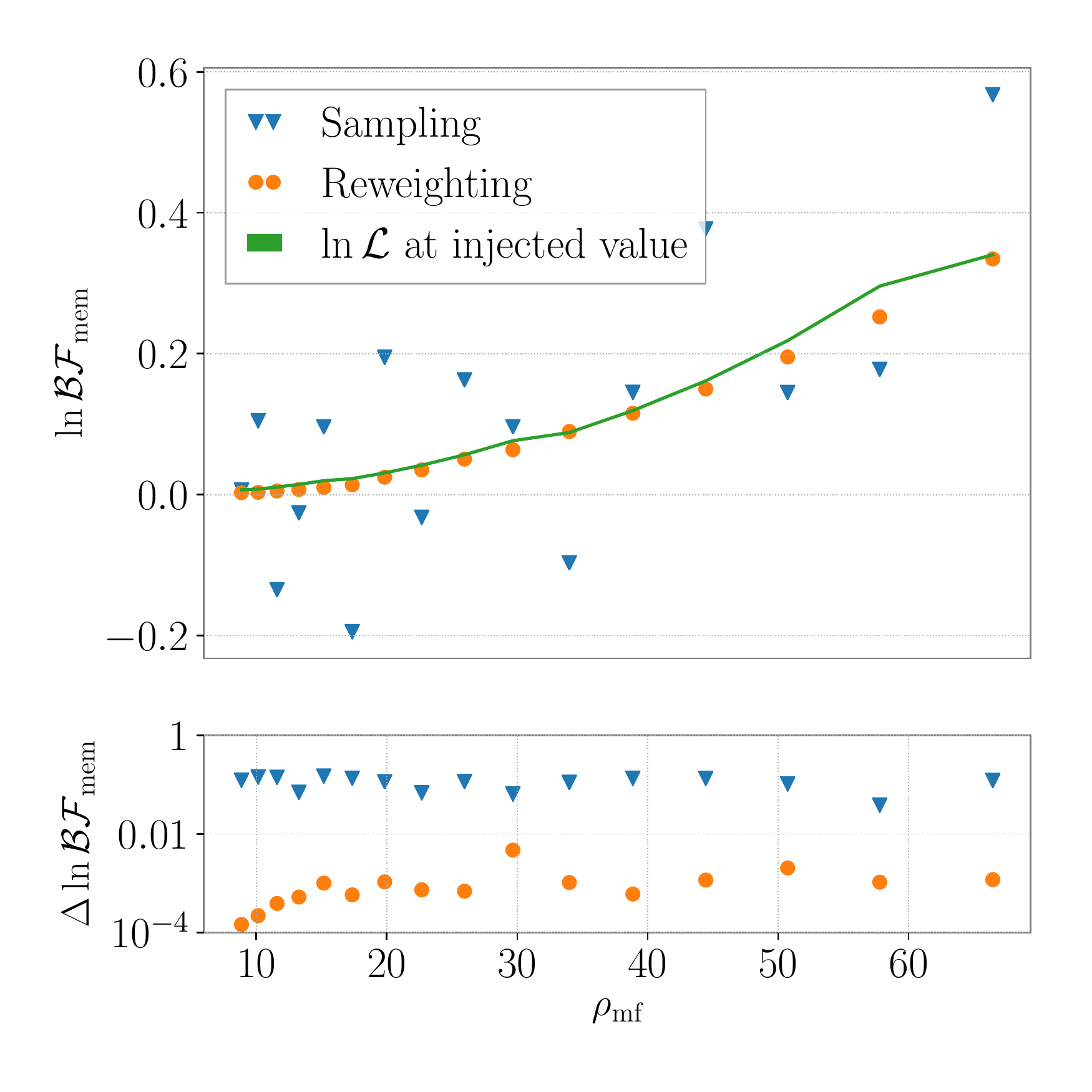}
    \caption{
    We compare the precision of calculating the Bayes Factors using different techniques for a set different signal-to-noise ratios $\rho_{\mathrm{mf}}$. 
    In the upper panel, we see that using the naive method of dividing the Bayes Factors from eight separate sampling runs (blue triangles), we see a much wider spread away from the fiducial line of likelihood ratios at the injected value (green curve) compared to the Bayes factors we obtained using reweighting (orange squares).
    In the lower panel, we see that using the reweighting method yields a statistical error that is about $\mathcal{O}(10^2)$ times smaller.
    }
    \label{fig:log_bfs_vs_snr}
\end{figure}

We use these software injections to compare reweighting to the naive method in which we carry out separate sampling runs with $\mathcal{L}_{\mathrm{osc}}$ and $\mathcal{L}_{\mathrm{full}}$.
Since this study is purely illustrative, we artificially break the $(\phi, \psi)$ degeneracy, by restricting the prior space by $\pm \pi/4$ around the injected values for $\phi$ and $\psi$.
By re-running the sampling algorithm eight times for each distance, we obtain an estimate of the uncertainty in the Bayes factor for both methods.
Finally, we also compare the estimates for the Bayes factor with the likelihood ratio at the injected parameter values, as this yields the Bayes factor one would obtain assuming perfect knowledge of the binary parameters.
The results are shown in Figure~\ref{fig:log_bfs_vs_snr}.
The upper panel \ref{fig:log_bfs_vs_snr} shows that reweighting is generally much better at recovering the Bayes factor whereas separately sampling both models can lead to significant sampling noise.
In the lower panel of Figure~\ref{fig:log_bfs_vs_snr} we display the stochastic error of both methods after eight runs.
This error ($\Delta\ln\BF$) is defined as the standard error of the sample mean of the eight $\ln\BF$s we obtained. 
Notably, the reweighting technique yields a reduction of about a factor $10^2$ in stochastic sampling noise.
Stochastic sampling noise vanishes with computation time $t$ as $\Delta\ln\BF \propto t^{-1/2}$~\cite{Chopin2010}, which implies that the $\sim 10^2$ improvement is equivalent to what would have been achieved by increasing the computation time by a factor of $\sim 10^4$.

\subsection{Analyzing real events}\label{sec:real_events}
The analysis of real events mostly follows the prescription in~\cite{HOM_REWEIGHTING}.
Initially, we perform inference with the \imr model to obtain a ``proposal'' posterior distribution.
Reweighting these posterior samples first with the \nrsur, a surrogate waveform model that includes modes $(\ell, m)$ up to $(5, 5)$~\cite{NRHYBSUR}, yields the Bayes factor for higher-order modes $\BF_{\mathrm{hom}}$, since \imr does not contain these modes.
Then reweighting with the full \nrsur plus memory model yields the combined higher-order mode plus memory Bayes factor $\BF_{\mathrm{hom+mem}}$.
The memory Bayes factor is
\begin{equation}
    \BF_{\mathrm{mem}} = \frac{\BF_{\mathrm{hom+mem}}}{\BF_{\mathrm{hom}}} \, 
\end{equation}

A final issue in the analysis is that \nrsur and \imr define the phase $\phi$ and time at coalescence $t_c$ differently, and there is no analytic way to map posterior samples between those two definitions.
Following~\cite{HOM_REWEIGHTING}, we map the posterior samples from \imr to \nrsur by maximizing the waveform overlap in terms of $\phi$ and $t_c$ between both models for each posterior sample.
The maximum overlap can be quickly found using common optimization techniques.
Furthermore, optimizing over the $(\phi,t_c)$ plane does not require us to evaluate the expensive \nrsur waveform at every step since these are not intrinsic parameters of the waveform. Instead, we produce the waveform once for each posterior sample and project it into the $(\phi,t_c)$ space as desired.
Results using this method analysing the gravitational-wave transient catalog are presented in Section~\ref{sec:results}.

\section{GWTC-1 Results}\label{sec:results}
\begin{figure}
    \centering
    \includegraphics[width=0.5\textwidth]{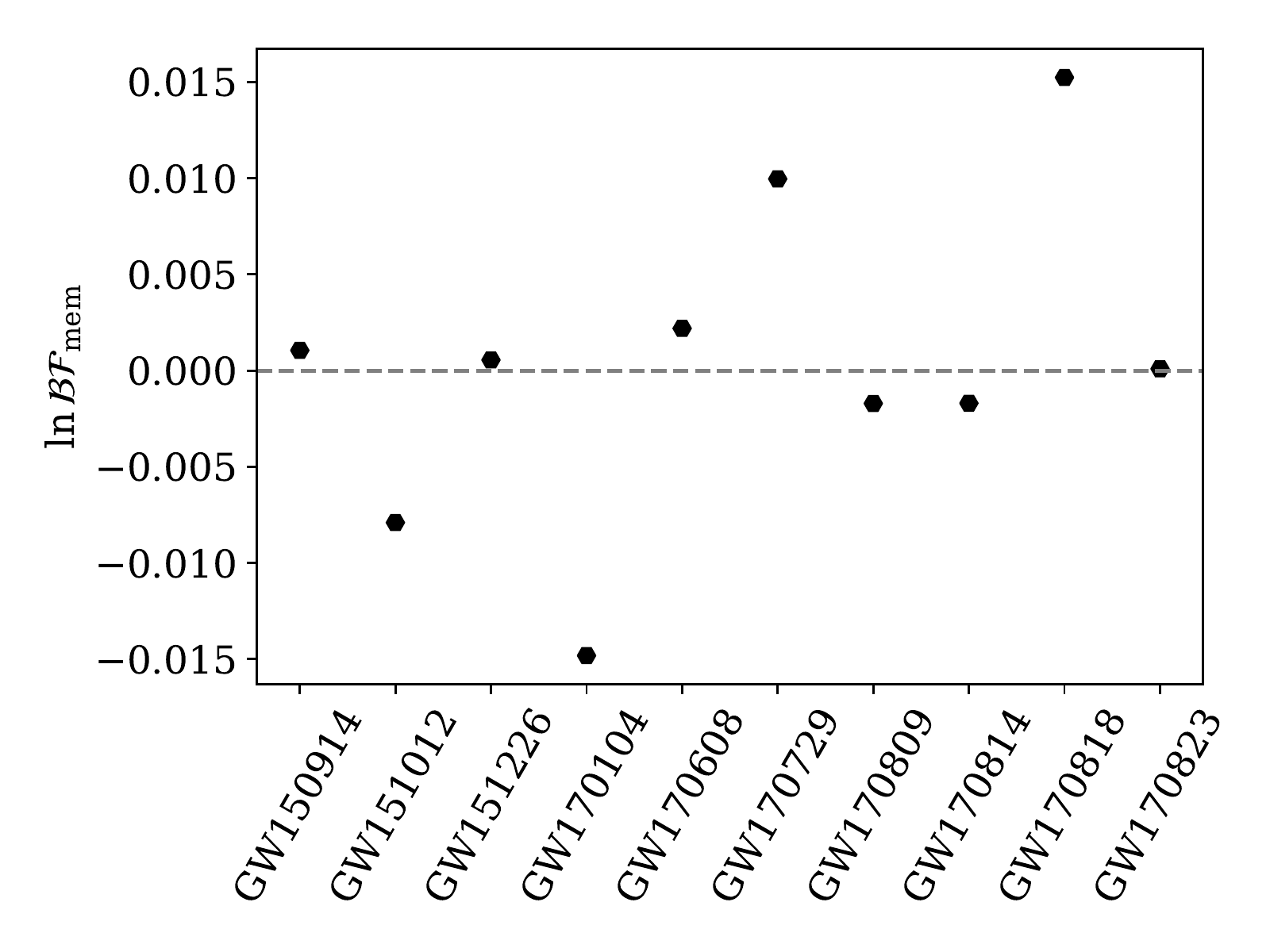}
    \caption{Memory Bayes factors obtained for the first two LIGO/Virgo observation runs.
    Overall, there is no significant evidence for or against the memory hypothesis.}
    \label{fig:gwtm1}
\end{figure}

We apply the reweighting technique on posterior samples of the first ten binary black hole mergers from the first two LIGO/Virgo observation runs.
The results are summarized in Figure~\ref{fig:gwtm1}.
The original posterior samples for the proposal run are the same as in~\cite{HOM_REWEIGHTING}.
The total $\ln \BF_{\mathrm{mem}}^{\mathrm{tot}} = 3.0\times 10^{-3}$ provides no significant support for or against the memory hypothesis.
However, this small Bayes factor is expected; we explore why in the subsequent section.
We see that even the loudest event in the catalog, GW150914 ($\rho_{\mathrm{mf}}\approx 26$), contributes only weak evidence in favour of the memory hypothesis.

\section{Population study}\label{sec:population}
We construct a simulated population of gravitational-wave events observed by the LIGO/Virgo detector network at design sensitivity so that we can estimate the number of required observations until we reach $\ln \BF \geq 8$.
We assume a power-law distribution both in primary mass and in mass ratio as outlined in~\cite{LIGO_RATES_AND_POPULATIONS}.
The mass distribution parameters are still poorly constrained given the low number of observations in the first two observing runs.
From the posterior distributions in~\cite{LIGO_RATES_AND_POPULATIONS} we choose parameters that correspond to the points of maximal posterior probability.
We choose minimum and maximum black hole masses $m_{\mathrm{min}} = 8\, M_{\odot}$ and $m_{\mathrm{max}} = 45\, M_{\odot}$ respectively, and use $\alpha = 1.5$ and $\beta = 3$ as spectral indexes for the primary mass and mass ratio distribution, respectively.

We assume an aligned spin prior distribution~\cite{RIFT_ALIGNED_SPIN_PRIOR}, with a maximal allowed spin magnitude of $a_{\mathrm{max}} = 0.5$. 
Higher spins are disfavoured observationally~\cite{LIGO_RATES_AND_POPULATIONS} and on theoretical grounds~\cite{FullerMa2019}.
At any rate, we do not expect the spin distribution to greatly affect the memory search because the absolute memory amplitude is mostly driven by the overall signal amplitude, which primarily depends on the masses and the luminosity distance of the source.
Spin only has an $\mathcal{O}(10\%)$ effect on the memory of a given binary.
The remaining extrinsic parameters (inclination, luminosity distance, sky position, time and phase at coalescence, polarisation angle) are chosen using standard priors.
We restrict the maximum luminosity distance to \SI{5000}{\mega\parsec} since more distant events are unlikely to be detected.

We randomly sample parameters from the distributions in intrinsic and extrinsic parameters.
However, the LIGO/Virgo detector network will only be able to actually detect a fraction of all occurring binary black hole mergers in the Universe.
We therefore only keep events with a matched filter signal-to-noise ratio greater than 12 in the network and/or greater than 8 in any single detector.
Otherwise, the event is considered to be undetected.

Following the steps outlined Section \ref{sec:real_events}, we obtain Bayes factors for each event.
In practice, this works reliably up to a matched filter signal-to-noise ratio $\rho_{\mathrm{mf}} \approx 32$, i.e. we recover the injected parameters and obtain an acceptable number of effective samples after reweighting~\cite{HOM_REWEIGHTING}.
At higher $\rho_{\mathrm{mf}}$, systematic differences between \imr and \nrsur can cause the inference runs to converge to non-overlapping regions in parameter space.
In those cases the reweighting technique using the \imr model becomes invalid if the posterior does not extend over the true value of the injected \nrsur data.
We resolve this issue by performing inference with the \nrsur model directly and then reweighting the posterior samples to the \nrsur plus memory model.
Since sampling with \nrsur is of far greater computational expense, we do not extend its use to the $\rho_{\mathrm{mf}} < 32$ events, which comprise  $92.5\%$ of all events in our population set.
Instead, we use the reweighting technique with \imr proposal distribution for these events.

We perform the analysis on a set of 2000 events and re-run inference until each combined posterior has at least 20 effective samples.
By requiring this number of effective samples, we ensure that the samples are reasonably closely converged to the injected value.
Otherwise, the weights would wildly diverge and the number of effective samples would hence always be close to unity.

\begin{figure}
    \centering
    \includegraphics[width=0.5\textwidth]{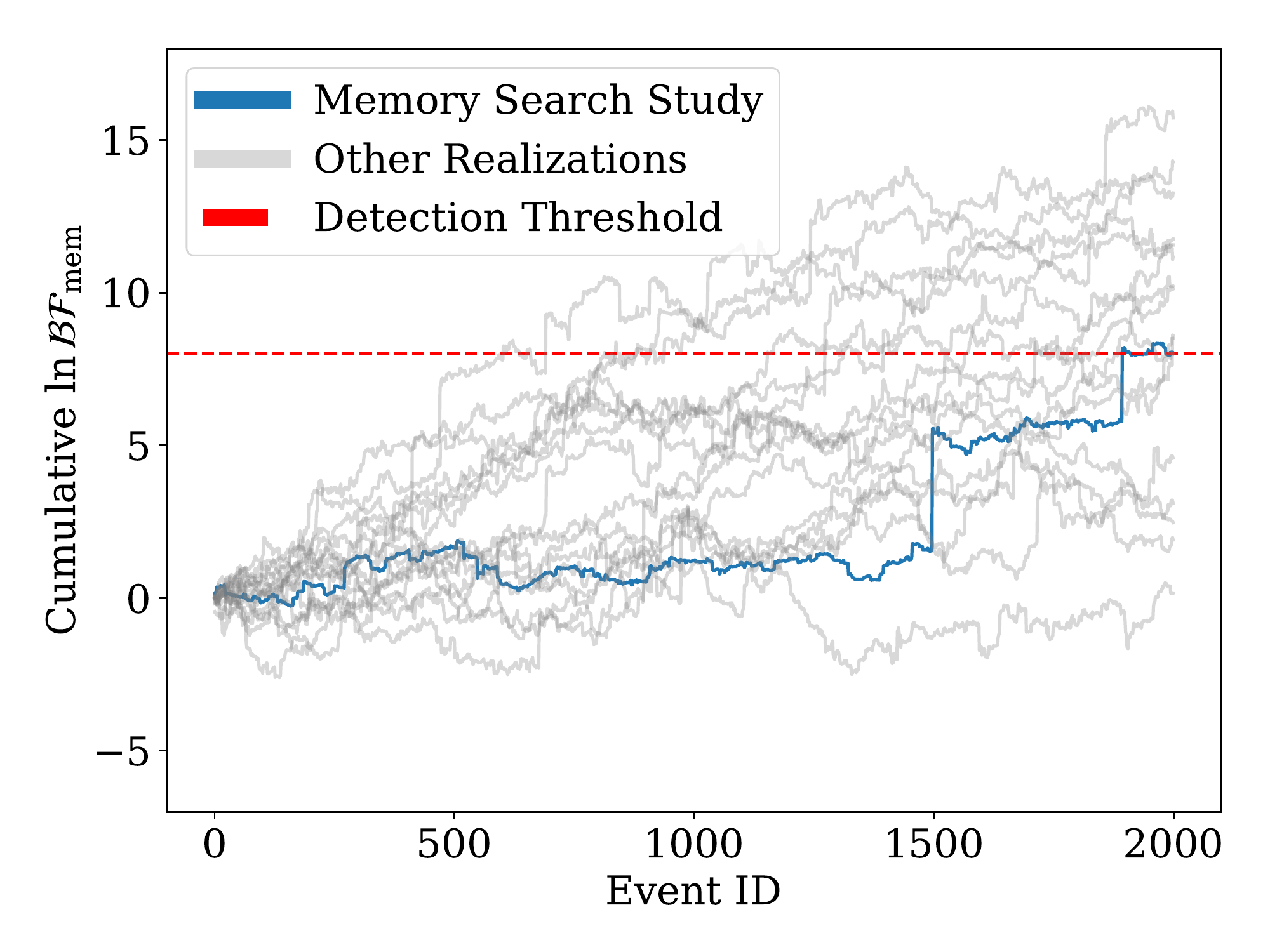}
    \caption{Cumulative memory Bayes factors obtained for a set of 2000 injections.
    The blue curve shows the recovered Bayes Factors using nested sampling and reweighting.
    The dashed red line displays the threshold for detection.
    The gray lines show a set of realizations using the the likelihood ratio at the injected parameters.}
    \label{fig:pop_study}
\end{figure}

We display the results of our population study in Figure~\ref{fig:pop_study} (blue curve). The population passes $\ln \BF > 8$ after about 2000 events.
We also simulate many more events for which we estimate the Bayes factor by using the likelihood ratio at the injected values (gray curves).
Using this much larger population, we estimate the required number of events to reach $\ln \BF \geq 8$ to be $1830^{+1730}_{-1100}$ at the $90\%$ confidence level.
Although this study likely overestimates the Bayes factors since it implicitly assumes that we can always break the $(\phi, \psi)$ degeneracy, we still consider this to be a good approximation since most support for memory comes from very few events with exceptionally high signal-to-noise ratios.
\section{Conclusion and Outlook}\label{sec:conclusion}
We have found a combined $\ln\BF = 0.003$ for the existence of memory in the gravitational waves from the ten binary black holes observed by LIGO/Virgo in their first two observing runs.
We have shown that we need $1830^{+1730}_{-1100}$ events to reach $\ln\BF = 8$, which can be considered to be a detection of memory~\cite{Lasky2016}.
This is likely to take place in the early days of A+/Virgo+, when observatories will be detecting ${\cal O}(10)$ events a day.
Adding KAGRA~\cite{KAGRA} and LIGO-India~\cite{LIGOIndia} to the network will further reduce the time until memory is detected.
Furthermore, reducing noise at low frequencies has also been shown to substantially decrease the number of detections required~\cite{LIGO_LOW_FREQ}, reducing the time to detection by a factor of $~3$.
Once memory is observed, it may be possible to use it to probe the nature of black holes and to look for physics beyond general relativity; see, e.g.,~\cite{Yang2018}.

We have shown how recent innovations, such as memory waveforms~\cite{Talbot2018}, and waveforms with higher-order modes enable us to know the sign of the memory, despite the computational challenges.
By introducing likelihood reweighting we reduce the stochastic sampling error by a factor of $\mathcal{O}(10^2)$, which is equivalent in terms of error reduction to an increase in sampling time by $\mathcal{O}(10^4)$.
Additionally, we show that by fine-tuning sampling parameters we can obtain confident measures of the Bayes factor within one week of computation time even if we have to use costly waveform models.

\section{Acknowledgements}\label{sec:acknowledgements}
We would like to thank Robert Wald for helpful comments. We would also like to thank Ethan Payne for kindly providing his posterior samples and helpful conversations.
This work is supported through Australian Research Council (ARC) Centre of Excellence CE170100004. PDL is supported through ARC Future Fellowship FT160100112 and ARC Discovery Project DP180103155.  ET is supported through ARC Future Fellowship FT150100281 and CE170100004.
This is LIGO Document No. DCC P1900346.

\bibliography{memory}

\end{document}